\documentclass[11pt]{article}
\usepackage{array}
\usepackage{amsmath}
\usepackage{amssymb}
\usepackage{graphicx}
\usepackage{amsthm}
\usepackage{latexsym}
\usepackage{algpseudocode}
\newtheorem{theorem}{Theorem}
\newtheorem{cor}{Corollary}
\newtheorem{definition}{Definition}
\newtheorem{prop}{Proposition}
\newtheorem{lemma}{Lemma}
\newtheorem{example}{Example}

\newcommand{\beq}{\begin{equation}}
\newcommand{\eeq}{\end{equation}}
\newcommand{\barr}{\left[\begin{array}}
\newcommand{\earr}{\end{array}\right]}

\newcommand{\bpf}{\emph{Proof}\/:}
\newcommand{\epf}{\hfill$\Box$}

\newcommand{\bi}{\begin{itemize}}
\newcommand{\ei}{\end{itemize}}
\newcommand{\bnum}{\begin{enumerate}}
\newcommand{\enum}{\end{enumerate}}
\newcommand{\bc}{\begin{center}}

\newcommand{\supp}{\ensuremath{\mbox{supp}\,}}
\begin{document}
\title{Generalized cofactors and decomposition of Boolean satisfiability problems}
\author{Madhav Desai\\madhav@ee.iitb.ac.in
\and 
Virendra Sule\\vrs@ee.iitb.ac.in
\and
Department of Electrical Engineering\\Indian Institute of Technology Bombay, India}
\maketitle
\date{}

\begin{abstract}
We propose an approach for decomposing Boolean satisfiability problems while extending recent results of \cite{sul2} on solving Boolean systems of equations. Developments in \cite{sul2} were aimed at the expansion of functions $f$ in orthonormal (ON) sets of base functions as a generalization of the Boole-Shannon expansion and  the derivation of the consistency condition for the equation $f=0$ in terms of the expansion co-efficients. In this paper, we further extend the Boole-Shannon expansion over an arbitrary set of base functions and derive the consistency condition for $f=1$. The generalization of the Boole-Shannon formula presented in this paper is in terms of \emph{cofactors} as co-efficients with respect to a set of CNFs called a \emph{base} which appear in a given Boolean CNF formula itself. This approach results in a novel parallel algorithm for decomposition of a CNF formula and computation of all satisfying assignments when they exist by using the given data set of CNFs itself as the base.
\end{abstract}

\section{Introduction}

The solution of systems of Boolean equations is one of the fundamental problems of Computer Science several of whose specialized formulations such as CNF satisfiability (SAT) have been under intense study since over the last fifty years \cite{hand,crah}. This is due to the vast applications of this problem in engineering, such as in the verification of software and computer hardware design, logic and artificial intelligence, analogy with several Computer Science problems \cite{bard, mezm} and in recent times to Biology \cite{kauf}. The 3-SAT problem is also well known as a model NP complete problem. In recent times there has been much interest in development of parallel solvers for the CNF SAT problem \cite{hamw}. For a scalable parallel solution it is important to have a strategy of decomposition of the CNF formula. A purpose of this paper is to arrive at such a strategy from the generalization of previous results on the well known Boole-Shannon expansion formula for Boolean functions. 

Essentially, in this paper we attempt to resolve the CNF-SAT problem and propose an algorithm which has its origin in a generalization of the well known Boole-Shannon formula which has been a basis of Boolean computation \cite{hand}. In \cite{sul1} a consistency condition for an equation $f(X)=0$ was first proved using an extension of this formula in terms of
an expansion of $f$ using a special basis set of what are called as \emph{orthonormal} (ON) functions. While this condition held true over general Boolean algebras the Boolean function $f$ had to be in the span of the ON set. In \cite{sul2} this condition was then further extended to cases when $f$ is not necessarily in the span of the ON set but then worked only for Boolean functions over the algebra $B_{0}=\{0,1\}$ which is of interest in applications.

In the present paper we first establish that the general Boole-Shannon expansion of functions relative to an ON set can in fact be proved even without orthonormality of the base set but with certain restrictions on $f$ relative to the chosen base set. This generalization actually stems from an observation that the expansion co-efficients relative to an ON set are just the \emph{cofactors} and satisfy the same algebraic constraints even for general base sets. The resulting consistency condition for the equation $f(X)=1$ in terms of expansion co-efficients of $f$ with respect to an arbitrary base set then generalizes the results of \cite{sul2}. We then propose an algorithm for CNF SAT using this condition and its consequent decomposition of the CNF formula. This decomposition gives rise to possibilities of efficient parallel solution to the SAT problem.

For the sake of comprehensibility we shall avoid introducing the background on Boolean algebra and functions and shall direct the reader to basic references \cite{rude, brow} and the previous paper \cite{sul2} which initiates the present line of investigation.

\section{Expansion formulas}

We present a brief background of the Boole-Shannon expansion and the generalized 
orthonormal expansion. We shall always be concerned with Boolean functions $f(X):B_{0}^{n}\longrightarrow B_{0}$ of $n$ variables denoted by $X$. The set of Boolean functions shall be denoted by $B_{0}(X)$. For an index $i$ the well known Boole-Shannon formula for $f$ is the following, which was in essence even known to Boole \cite{bool} while Shannon re-established its role in the 20th century,
\[
f(X)=x_{i}f(x_{i}=1)+x_{i}'f(x_{i}=0)
\]
This formula has a generalization in terms of expansion with respect to ON functions. A set of functions $\{\phi_{1}(X),\ldots,\phi_{m}(X)\}$ is said to be orthogonal if, for $i \neq j$, $\phi_{i}\phi_{j}=0$ the zero function. The set is orthonormal (ON), when in addition
\[
\sum_{i=1}^{m}\phi_{i}(X)=1
\]
It is easily proved and shown in \cite{brow} that any Boolean function $f$ has an ON expansion
\begin{equation}\label{ONexp}
f(X)=\sum_{i=1}^{m}\alpha_{i}(X)\phi_{i}(X)
\end{equation}
where the expansion co-efficient functions $\alpha_{i}(X)$ satisfy the interval
\[
\alpha_{i}\in [f\phi_{i},f+\phi_{i}']
\]
The central theme of the previous paper \cite{sul2} was to determine a consistency condition for the equation $f(X)=0$ in terms of the co-efficients $\alpha_{i}$. This paper shall report further generalizations and results.

\subsection{Expansion in ON terms}

The ON expansion shown in Eq. (\ref{ONexp}) has the well known special case in which the ON functions are ON terms (products). If $T=\{t_{1},\ldots,t_{m}\}$ are ON terms in variables $X$ then any Boolean function $f(X)$ has the dual expansions \cite[Theorem 3.15.1]{brow}
\[
\begin{array}{rcl}
f(X) & = & \sum_{i=1}^{m}(f/t_{i})(X)t_{i}(X)\\
f(X) & = & \prod_{i=1}^{m}[(f/t_{i})(X)+t_{i}(X)']
\end{array}
\] 
where $(f/t)(X)$ for any term $t(X)$ in $n$ variables $X$ is called the \emph{quotient} which is defined in the literature \cite{brow, koha} as the function $f(X)$ along with the constraints on variables defined by the equation
\[
t(X)=1
\]
Clearly for a term $t$ the constraint $t=1$ results in \emph{partial assignments} of variables in $X$ which we shall denote as $q(t)$. For instance if $X=\{x_{1},x_{2},x_{3}\}$ then for $t=x_{1}x_{3}'$, $q(t)=\{x_{1}=1,x_{3}=0\}$. Hence $f/t=f(1,x_{2},0)$ is a function of $x_{2}$. The definition of quotient thus does not extend to the case when $t$ is replaced by a sum of terms.

\subsection{Generalization to cofactors}

An identical concept to the quotient $f/t$ talked about in literature \cite{koha} is the concept of \emph{cofactor} of a function $f$ relative to a term $t$ denoted as $f_{t}$. However it is defined again as the function $f$ considered along with the constraint equation $t=1$. We shall thus extend this definition for the general case when the term $t$ is another function.

\begin{definition}\label{dfn:cofactor}\emph{
Let $f(X)$, $g(X)\neq 0$ be Boolean functions in $n$ variables $X$ in $B_{0}(X)$. Then a set of \emph{cofactors} of $f$ relative to $g$ denoted $\Xi(f,g)$ is the set of all Boolean functions whose restriction on the set $\supp (g)$\footnote{$\supp g=\{x\in B_{0}^{n}|g(x)=1\}$} is identical to that of the restriction of $f$ on $\supp (g)$.
}
\end{definition}

The cofactors are thus characterized as follows

\begin{lemma}\emph{
The set of cofactors $\Xi(f,g)$ is the set of functions $\alpha (X)$ which satisfy the equation
\beq\label{eq:cofactordefn}
\alpha(X)g(X)=f(X)g(X)
\eeq
Hence $\Xi(f,g)$ is given by the interval
\[
[f(X)g(X),\,f(X)+g(X)']
\]
and in parametrized form
\[
\Xi(f,g)=\{fg+pg', p\in B_{0}(X)\}
\]
}
\end{lemma}

\bpf
If any function $\alpha(X)$ is in $\Xi(f,g)$ then $\alpha g=fg$. Conversely if $\alpha$ is a solution of this equation in $B_{0}(X)$ then $\alpha(x)=f(x)$ for $x$ in $\supp (g)$. The interval shown is the set of all solutions $\alpha$ of this equation. Hence this interval is the set of all cofactors. Every element of this interval is of the form $fg+pg'$ for an arbitrary function $p$.
\epf

The set $\Xi(f,g)$ has following properties w.r.t. Boolean operations.

\begin{prop}\emph{ Let $\alpha$, $\beta$ belong to $\Xi(f,g)$ then
\begin{enumerate}
\item $\alpha+\beta$ belongs to $\Xi(f,g)$.
\item $(\alpha)(\beta)$ belongs to $\Xi(f,g)$.
\item $\alpha'$ belongs to $\Xi(f',g)$ given by the interval $[f'g,f'+g']$ and has the general form
\[
\alpha'=f'g+pg'
\]
for an arbitrary function $p$.
\end{enumerate}
}
\end{prop}

Note that the preceding proposition implies that $| \Xi (f,g) | = | \Xi (f',g)|$.

\subsection{Representation in cofactors: beyond orthonormal expansion}

Now consider the ON set $g,g'$ and let
\[
f=(\alpha)g+(\beta)g'
\]
be an ON expansion of $f$ in this set. Then it is well known that $\alpha$, $\beta$ satisfy the equations
\[
\begin{array}{rcl}
(\alpha)g & = & fg\\
(\beta)g' & = & fg'
\end{array}
\]
Hence $\alpha$ belongs to $\Xi(f,g)$ and $\beta$ belongs to $\Xi(f,g')$. Hence these are just cofactors with respect to an ON set $\{g,g'\}$. From this observation we can explore generalization of the expansion formulas as follows.

\subsubsection{Boole-Shannon expansion}

The well known Boole-Shannon expansion formula for Boolean functions $f:B_{0}^{n}\rightarrow B_{0}$ in $n$-variables $X$ is
\[
f=x_{i}f(x_{i}=1)+x_{i}'f(x_{i}=0)
\]
This is generalized by the ON expansion (\ref{ONexp}) above. However in terms of the cofactors of $f$ with respect to $g, g'$, a further generalization of Boole-Shannon expansion formula is as established above

\begin{prop}
\beq\label{Shannon}
f=(\alpha)g+(\beta)g'
\eeq
for any $\alpha \in \Xi(f,g)$ and $\beta \in \Xi(f,g')$.
\end{prop}

The above expression replaces the ON set $\{x,x'\}$ in one of the variables to the ON set $\{g,g'\}$. The concept of cofactors however allows a further generalization even when the base set is not ON. This is developed in the next subsection.

\subsubsection{Beyond orthonormality}

Consider the set $\{g,h\}$ of two Boolean functions. Let $f$ be a given Boolean function whose cofactors w.r.t. this set are $\Xi(f,g)$ and $\Xi(f,h)$. In analogy with the ON expansion formula for $f$ in terms of $g,g'$, we ask, what should be conditions on $g,h$ such that the function
\[
F=(\alpha)g+(\beta)h
\]
is equal to $f$ for any choice of $\alpha \in \Xi(f,g)$ and $\beta \in \Xi(f,h)$?

Since for any $x$ in $B_{0}^{n}$ and the fact that $\alpha,\beta$ are cofactors as above
\[
F(x)=g(x)f(x)+h(x)f(x)=(g(x)+h(x))f(x)
\]
Hence it follows that $F=f$ iff
\[
f\leq g+h
\]
we can thus look for a general statement as in

\begin{theorem}
\emph{Let $G=\{g_{1},\ldots,g_{m}\}$, $g_{i}\neq 0$ for any $i$, be a set of Boolean functions whose sum is
\[
g=\sum_{i=1}^{m}g_{i}
\] 
Then for any Boolean function $f\leq g$
\beq\label{cofac}
f=\sum_{i=1}^{m}\alpha_{i}g_{i}
\eeq
for any $\alpha_{i}\in \Xi(f,g_{i})$.
}
\end{theorem}

\bpf
Let
\[
F=\sum_{i=1}^{m}\alpha_{i}g_{i}
\]
where $\alpha_{i}$ in $\Xi(f,g_{i})$ be chosen arbitrarily. Then for any $x$ in $B_{0}^n$,
\[
F(x)=\sum_{i}f(x)g_{i}(x)=f(x)(\sum_{i}g_{i}(x))=f(x)g(x)
\]
Hence $F=fg$. Since $f\leq g$ it follows that $F=f$.
\epf

We call this last expansion (\ref{cofac}) as \emph{generalized cofactor expansion} of $f$ relative to $G$ and it is quite apparent that this generalizes the ON expansion formula (\ref{ONexp}). This expansion is then a vast generalization of the Boole-Shannon formula going beyond orthonormality.

\begin{example}\emph{
Consider $f=x_{1}'x_{2}+x_{2}x_{3}+x_{1}x_{3}'$ and $C=(x_{1}'+x_{3})$ then
\[
fC=x_{1}'x_{2}+x_{2}x_{3}=x_{2}C=x_{1}'x_{2}+x_{2}x_{3}
\]
which is the minimal co-factor in $\Xi(f,C)$ while
\[
f+C'=f
\]
and hence the maximal cofactor is $f$ itself. A general function in $\Xi(f,C)$ is
\[
fC+pC'=x_{2}(x_{1}'+x_{3})+(p)(x_{1}x_{3}')
\]
where $p$ is an arbitrary Boolean function in the three variables.
}
\end{example}

\subsection{Algebra in terms of cofactors}
As in the case of the ON expansion of functions, various computations and operations in the Boolean algebra of functions can be expressed in terms of co-factors since the equation defining the co-factors is the same as that of the equation determining the expansion co-efficients of an ON expansion. For the purpose of completeness these identities are presented below.

\begin{prop}\emph{
Let $G=\{g_{1},\ldots,g_{m}\}$ be a base set as in the above theorem and $f,h$ be Boolean functions with cofactors relative to $G$ denoted as $\alpha_{i} \in \Xi(f,g_{i})$ and $\beta_{i} \in \Xi(h,g_{i})$. Then 
\[
\begin{array}{rcl}
\alpha_{i}+\beta_{i} & \in & \Xi(f+h,g_{i})\\
(\alpha)(\beta)  & \in & \Xi(fh, g_{i})\\
\alpha_{i}'   & \in & \Xi(f', g_{i})\\
\alpha_{i}\oplus\beta_{i} & \in & \Xi(f\oplus h, g_{i})
\end{array}
\]
}
\end{prop}

\bpf
The proofs follow easily from the characterization of $\Xi(f,g_{i})$ and $\Xi(h,g_{i})$.
\epf

In fact the above relations prove that the following identities hold similar to the ON expansion formulas in \cite[section 3.14.1]{brow}

\begin{eqnarray}\label{expidentties}
f(X)+h(X) & = & \sum_{g_{i}\in G}(\alpha_{i}(X)+\beta_{i}(X))g_{i}(X)\\
f(X)h(X) & = & \sum_{g_{i}\in G}(\alpha_{i}(X)\beta_{i}(X))g_{i}(X)\\
f(X)' & = & \sum_{g_{i}\in G}\alpha_{i}(X)'g_{i}(X)\\
f(X)\oplus g(X) & = & \sum_{g_{i}\in G}(\alpha_{i}(X)\oplus\beta_{i}(X))g_{i}(X)
\end{eqnarray}

Composition of Boolean functions can also be expressed in terms of above expansion formulas as follows. The simple proof is omitted. Consider the base set $G$ as above and let $h_{i}(X)$ for $i=1,\ldots,n$, $f(X)$ be Boolean functions in $n$-variables and
\[
h_{i}(X)=\sum_{j}\beta_{ij}(X)g_{j}(X)
\]
be expansions of $h_{i}(X)$ in the set $G$ as above. Then
\[
f(h_{1}(X),\ldots,h_{n}(X))=\sum_{j}\alpha_{j}(\beta_{1j}(X),\ldots,\beta_{nj}(X))\phi_{j}(X)
\]

\section{Consistency conditions from the generalized cofactor expansion}

As shown in \cite{sul2} the ON representation despite its non-unique coefficients is useful for solving Boolean equations and from a theoretical standpoint even for deriving consistency conditions on $f=0$ when an arbitrary ON expansion of $f$ is known. In fact it is shown in \cite{sul2} that the problem of satisfiability of systems of Boolean equations and that of determining all solutions of such systems can be solved by ON expansion of component functions without the need for forming a single equation from the system.

In view of the above generalization of the ON expansion formula in terms of co-factors it is instructive to explore as far as feasible, the extension of the program followed in \cite{sul2} in terms of cofactors associated with systems of Boolean equations. We begin this development with a preliminary result.

\begin{theorem}\label{pr:con}\emph{Let $G=\{g_{1},\ldots,g_{m}\}$ be a set of non-zero Boolean functions
\[
g=\sum_{i=1}^{m}g_{i}
\]
and $f\leq g$. Then $f=1$ is consistent iff there exists an index $i\in [1,m]$ such that the system
\[
\alpha_{i}=1,\; g_{i}=1
\]
is consistent for any cofactor $\alpha_{i} \in \Xi(f,g_{i})$.
}
\end{theorem}

\bpf
If $f(x)=1$ for some $x$ then it follows from the expansion (\ref{cofac}) that for some $i$, $\alpha_{i}(x)g_{i}(x)=1$. Hence $x$ is in $\supp g$ and $\alpha_{i}(x)=1$ and thus the system is consistent and this is independent of the choice of $\alpha_{i}$ in $\Xi(f,g_{i})$. Conversely if $f=1$ is not consistent then $f(x)=0$ for all $x$ in $B_{0}^{n}$. Hence for any $x$, any $i$ and a choice of cofactor $\alpha_{i}$, $\alpha_{i}(x)g_{i}(x)=0$. This shows that the system $\alpha_{i}=1$, $g_{i}=1$ is not consistent.
\epf

In terms of an ON base set we can obtain the following result. Let $\Phi=\{\phi_{1},\ldots,\phi_{m}\}$ be an ON set of non-zero functions. Then for any Boolean function $f$ we have the cofactor based expansion of the form (\ref{cofac})
\[
f=\sum_{i=1}^{m}\alpha_{i}\phi_{i}
\]
where $\alpha_{i} \in \Xi(f,\phi_{i})$.

\begin{cor}\emph{The equation $f(X)=1$ is consistent iff there is exactly one index $i$ such that the system 
\[
\alpha_{i}(X)=1,\;\phi_{i}(X)=1
\]
is consistent for an arbitrary choice of co-factor in the above expansion.
}
\end{cor}

\bpf
An ON expansion of $f$ in the base set $\Phi$ based on cofactors, of the form
\[
f=\sum_{i=1}^{m}\alpha_{i}\phi_{i}
\]
exists as sum of functions in $\Phi$ is $1$. Since $\Phi$ is ON and the functions are non-zero, if the equation is consistent and $f(x)=1$ for some $x$. Since there is exactly one index $i$ for which the function $\phi_{i}(x)=1$ for any $x$ while all others vanish at this $x$ it follows that the given system is consistent for this $i$ irrespective of the co-factor $\alpha_{i}$. This proves necessity.

Conversely if the given system is consistent for some $i$ and an $x$ is a solution, then as $\phi_{j}$ are orthogonal to $\phi_{i}$ we get $\phi_{i}(x)=1$ and $\phi_{j}(x)=0$ for $j\neq i$. Since $\alpha_{i}=f\phi_{i}+p\phi_{i}'$, this implies $f(x)=1$ for this $x$ which proves sufficiency.
\epf

\section{Decomposition of CNF formulas}
Consistency results in the form shown above should be useful for the
solution of a variety of problems of Boolean satisfiability. In particular we explore decomposing a CNF formula for satisfiability. Let
\[
f=\prod_{i=1}^{m}C_{i}
\]
be a CNF formula expressed as a product of CNFs. Then the condition
\[
f\leq\sum_{i=1}^{m}C_{i}
\]
holds. Hence we have a representation of $f$ as in (\ref{cofac})
\[
f=\sum_{i=1}^{m}\alpha_{i}C_{i}
\]
where $\alpha_{i}$ belong to $\Xi(f,C_{i})$. However we show next that the actual computation of or choice of $\alpha_{i}$ for each $C_{i}$ can be avoided while theorem \ref{pr:con} can be still utilized for exploring satisfiability.

\subsection{Partial assignments and their substituion}
To explain the application of theorem \ref{pr:con} to the CNF SAT problem we need to introduce the notion of \emph{partial assignment} defined by a clause and notation for substituion of partial assignments in functions. Consider a clause,
\[
C=x_{i_{1}}^{a_{1}}+\ldots+x_{i_{k}}^{a_{k}}
\]
with $a_{i}=0,1$ defined by the rule $x^{1}=x$, $x^{0}=x'$. Then $C=1$ for any of the assignments of variables in the set $s(C)=\{x_{i_{j}}=a_{j}\}$. This set is called SAT set of $C$. Hence the set denoted $q(C)$ called the set of partial assignments defined by $C$, which consists of all non empty subsets of $s(C)$, is the set of assignments of arguments in $C$ which lead to evaluate $C=1$. For example the CNF
\[
C=x_{1}'+x_{2}+x_{3}'
\]
has the SAT set $s(C)=\{x_{1}=0,x_{2}=1,x_{3}=0\}$. Hence the set $q(C)$ of all partial assignments consists of subsets denoted by columns of the table
\[
\begin{array}{|c|c|c|c|c|c|c|c|}
\hline
x_{1} & 0 & & & 0 & 0 & & 0 \\
\hline
x_{2} &  & 1 & & 1 & & 1 & 1\\
\hline
x_{3} & & & 1 & & 1 & 1 & 1 \\
\hline
\end{array}
\]
For a CNF containing $k$ terms therefore there are
\[
\sum_{r=1}^{k}\binom{k}{r}
\]
elements in $q(C)$.

If $C$ is one of the clauses in a product formula $f=C_{1}C_{2}\ldots C_{m}$ then for any fixed $i$, substitution of each partial assignment from $q(C_{i})$ results into reduced CNF formulas. 

\subsection{Satisfiability by decomposition}
By a reformulation of theorem \ref{pr:con} we get a decomposition of a CNF formula. Let
\[
f=\prod_{1=1}^{m}C_{i}
\]
be a CNF formula.

\begin{cor}\emph{$f$ is SAT (i.e. $f=1$ is consistent) iff there exists an index $i$ and a partial assignment $q$ in $q(C_{i})$ such that the CNF formula
\[
f(q)
\]
after partial assignment $q$ is substituted in $f$ is SAT.
}
\end{cor}

\bpf
Let $\alpha_{i}$ be a cofactor in $\Xi(f,C_{i})$. Then by theorem \ref{pr:con}, $f$ is SAT iff for some $i$ there is $q$ in $q(C_{i})$ such that $\alpha_{i}(q)=1$ is satisfiable. Since partial assignment of variables at this $q$ gives $\alpha_{i}(q,z)=f(q,z)$ for other unassigned variables $z$, it follows that the condition is equivalent to satisfiability of $f(q,z)$.
\epf

For each partial assignment $q$ there is the CNF formula $f(q)$ whose satisfiability needs to be tested to determine satisfiability of $f$. However $f(q)$ is a reduced formula compared to $f$ and for different $q$ these are independent SAT problems. This gives rise to a decomposition of the original SAT problem into reduced problems which can be solved in parallel. An algorithm based on this decomposition is developed in the next section. We conclude with an illustrative example.

\begin{example}\emph{Consider the CNF formula
\[
f=C_{1}C_{2}C_{3}C_{4}=(x'+y+w)(y'+z+w')(x+z+w')(x+z'+w')
\]
The SAT sets of the CNF factors are as follows
\[
\begin{array}{rcl}
s(C_{1}) & = & \{x=0,y=1,w=1\}\\
s(C_{2}) & = & \{y=0,z=1,w=0\}\\
s(C_{3}) & = & \{x=1,z=1,w=0\}\\
s(C_{4}) & = & \{x=1,z=0,w=0\}
\end{array}
\]
The set $q(C_{1})$ is thus shown by the indexed columns of the table
\[
\begin{array}{|c|c|c|c|c|c|c|c|}
\hline
  & 1 & 2 & 3 & 4 & 5 & 6 & 7\\
\hline
x & 0 & & & 0 & 0 & & 0\\
\hline
y & & 1 & & 1 & & 1 & 1\\
\hline
w & & & 1 & & 1 & 1 & 1\\
\hline
\end{array}
\]
Reduced CNFs at various partial assignments in $q(C_{1})$ are shown below indexed by the columns in the above table
\[
\begin{array}{rl}
1 & (y'+z+w)(z+w')(z'+w')\\
2 & (z+w')(x+z+w')(x+z'+w')\\
3 & (y'+z)(x+z)(x+z')\\
4 & (z+w')(z'+w')\\
5 & (y'+z)(z)(z')\\
6 & (z)(x+z)(x+z')\\
7 & (z)(z')
\end{array}
\]
Similarly reduced CNFs at $q(C_{i})$ for $i=2,3,4$ can be computed.
}
\end{example}

\subsection{All solutions of a CNF formula}

When a CNF formula denoted $F(C)$ which is the product of all CNFs in a set $C$ of CNFs is satisfiable, the set $L(C)$ of all solutions of $F=1$ (also called the SAT solutions of $C$) is the intersection of all partial assignments satisfying CNFs in $C$. Representing such a set of solutions efficiently is a problem of developing a suitable data structure. We consider next the problem of decomposing a formula $F$ in terms of smaller formulas using the decomposition facilitated by the above theory. The problem of returning all solutions of a formula is yet another problem which shall be addressed elsewhere. 

\subsection{Algorithm for CNF SAT}

The idea of the algorithm to be described now follows from the example of the last section. The algorithm identifies parallel computations due to the inherent decomposition available and is built using following functions. We consider $C$ a set of given CNFs, $F(C)$ the formula which is the product of all CNFs in $C$. The algorithm utilizes following functions
\begin{enumerate}
\item The function \textbf{Decompose()} which decomposes a given data set $C$ of CNFs in terms of sets of smaller size from which all solutions can be computed by solving all solutions for each smaller set.
\item A function \textbf{Varlist()} extracts and indexes the variable list inside a set $C$.
\item $n_{0}$ the smallest number of variables in $C$ at which further decomposition is not utilized for computation of solutions.
\item Function \textbf{Partial-assignments()} returns all partial assignments satisfying a set $C$.
\end{enumerate}

\begin{algorithmic}
\Function {Decompose}{$C$,$X$,$n_{0}$}
\State \textbf{Data} $C$ Set of CNFs, $X$ variable list in $C$, $n_{0}$ number. 
\State \textbf{Output} CNF sets with number of variables $\leq n_{0}$ whose union of solutions is equal to the set of all SAT solutions of $C$.
\Loop \% The main Loop
\While {$|X|> n_{0}$}
\State Choose a subset $X_{1}\subset X$ of variables, $|X_{1}|\leq n_{0}$.
\State Partition $C=C_{1}\bigsqcup C_{2}\bigsqcup C_{3}$, $X=X_{1}\bigsqcup X_{2}$
\State \%$C_{1}$ set of clauses containing only $X_{1}$ variables.
\State \%$C_{2}$ set of clauses containing elements of both $X_{1}$ and $X_{2}$.
\State \%$C_{3}$ set of clauses containing only $X_{2}$ variables.
\State Compute all partial assignments $L=\mbox{\textbf{Partial-assignment}}(C_{1})$.
\If {$L=\emptyset$}
\State Unsatisfiable formula.
\State Exit with output \texttt{UN-SAT}
\Else
\State Gather all partial assignments at a memory element. 
\State Identify indices of variables assigned.
\EndIf
\For {$q_{i}\in L$}
\State Compute $C_{2}(q_{i})$ 
\State Send $C_{2}(q_{i})$ to independent processing elements $PE(i)$.
\State Send $C_{3}$ to an independent processing element $PE(|L|+1)$.
\EndFor
\EndWhile
\State Return all satisfying assignments of $C$.
\EndLoop
\State Exit the algorithm.

\For {each $PE(i)$}
\State $C\gets C_{2}(q_{i})$ 
\State $X\gets \mbox{\textbf{Varlist}}(C)$
\State Repeat the main loop at the $PE(i)$
\EndFor

\For {the processing element $PE(|L|+1)$}
\State $C\gets C_{3}$
\State $X\gets \mbox{\textbf{Varlist}}(C)$
\State Repeat the main loop
\EndFor
\EndFunction
\end{algorithmic}

The above function decomposes the original CNF formula into independent sub-formulas of variable sizes less than $n_{0}$. The set of all SAT solutions of the original formula is then the union of all solutions of all resultant sub-formulas. The set of all SAT solutions of the original formula is obtained by patching the partial assignments obtained during decomposition (gathered at the memory node) and all solutions of the resultant sub-formulas obtained at independent processing elements. The original formula is un-SAT iff all the final sub-formulas are un-SAT. This last loop can be stated as follows:

\begin{algorithmic}
\Function{Gatherallsolutions}{$C$}
\For {$C$ at each $PE$}
\State Compute all SAT assignments $s(C)$ of $C$.
\State gather $s(C)$ at the memory node and patch the solution to the existing assignments.
\EndFor
\EndFunction
\end{algorithmic}

As stated above we do not discuss the actual algorithm for this function, it is assumed that such a function which computes all solutions of a small enough CNF set $C$ is available. 

\subsection{Computing all partial assignments}
For a CNF data $C$ in sufficiently small number of variables i.e. $<n_{0}$ computation of the set of all assignments $L$ satisfying $C$ is required to be carried out in the algorithm. This is a well known $\sharp P$ complete problem hence this loop can only be executed for a sufficiently small size of $C$. This problem of enumerating all assignments of a CNF formula efficiently thus requires detailed data structure and shall be discussed elsewhere.

\subsection{Estimate of average computation time}
The  algorithm described above essentially only shows the decomposition of a complex CNF formula to smaller formulas with number of variables at most $n_{0}$. The final computation of all solutions is accomplished by directly solving the decomposed formulas in parallel. The computation of time complexity can thus be carried out from following estimates.
\begin{enumerate}
\item $T_{0}$: average time complexity of generating all solutions of CNF formula of variable size at most $n_{0}$.
\item $d$: maximum number of decompositions required for a formula with number of variables $n$, called depth,
\[
\mbox{depth}\;d=N\div n_{0}
\]
the quotient of $N$ when divided by $n_{0}$.
\item Number of variables in the final decomposition is at most
\[
r_{0}=N\mbox{mod}\;n_{0}
\]
\item Average time required for substitution of solutions for decomposition $S_{0}$. 
\end{enumerate}

Hence the average time required for solving the complete problem by decomposition is
\[
T=T_{0}^{d}+d\times S_{0}
\]
Hence the $T_{0}^{d}$ term dominates the computation time.

\section{Conclusions}
We develop a generalization of the well known Boole-Shannon representation of Boolean functions in terms of cofactors relative to a base set of functions. For a CNF formula the base set can be chosen from the CNF factors themselves. This leads to a decomposition of the CNF formula which is shown to be useful for making a parallel solver for all satisfying assignments of the formula. However the algorithm developed on the basis of this representation is less surprising since it essentially shows substitution of partial solutions for decomposition. The intermediate step of computing all assignments of a smaller problem is a well known $\sharp P$ complete problem hence $n_{0}$ should necessarily be small in practice. Many heuristic improvements can be developed centered on this algorithm which will be discussed elsewhere.

\begin{center}
Acknowledgements

Second author greatfully acknowledges support from the project grant number 11SG010 of IRCC of IIT Bombay.
\end{center}


\begin{thebibliography}{xxxx}
\bibitem{bool} George Boole. An Investigation of the Laws of thought. Walton, London, 1854.
\bibitem{brow} F.\ M.\ Brown. Boolean reasoning. The logic of Boolean equations. Dover, 2006.
\bibitem{rude} Sergiu Rudeanu. Boolean functions and equations. North Holland, Amsterdam, 1974.
\bibitem{hand} A.\ Biere, M.\ Heule, Hans van Maaren, T.\ Walsh (Eds). Handbook of Satisfiability. IOS Press, 2009.
\bibitem{crah} Yves Crama and Peter Hammer. Boolean functions. Theory, algorithms and applications. Encyclopedia of Mathematics and its applications, vol.142. Cambridge, 2011.
\bibitem{bard} Gregory Bard. Algebraic cryptanalysis. Springer 2009.
\bibitem{mezm} Marc Mezard and Andrea Montanari. Information, Physics and Computation. Oxford University Press, 2009.
\bibitem{kauf} Stuart Kauffman. Origins of order, self organization and selection in evolution. Oxford University Press, 1993.
\bibitem{hamw} Youssef Hammadi and C.\ M.\ Wintersteiger. Seven challenges in parallel SAT solving. Challenge paper AAAI 2012 Sub-Area spotlights track. Association of Advancement of Artificial Intelligence.
\bibitem{koha} Kohavi and Jha, Switching and automata theory, Cambridge 2008.
\bibitem{sul1} Virendra Sule, Generalization of Boole-Shannon expansion, consistency of Boolean equations and elimination by orthonormal expansion, arXiv.org/cs.CC/1306.2484v3, December 6, 2013.
\bibitem{sul2} Virendra Sule, An algorithm for Boolean satisfiability based on generalized orthonormal expansion, arXiv.org/cs.DS/1406.4712v3, July 16, 2014.
\end{thebibliography}
\end{document}